\documentclass[%
 reprint,
superscriptaddress,
 amsmath,amssymb,
 aps,
]{revtex4-2}

\usepackage{graphicx}
\usepackage{dcolumn}
\usepackage{bm}
\usepackage{comment}
\usepackage{siunitx}

\begin{document}

\preprint{APS/123-QED}

\def\be{\begin{equation}}
\def\ee{\end{equation}}

\title{Enhanced on-chip frequency measurement using weak value amplification}

\author{John Steinmetz}
\email{jsteinm3@ur.rochester.edu}
\affiliation{Department of Physics and Astronomy, University of Rochester, Rochester, NY 14627, USA}
\affiliation{Center for Coherence and Quantum Optics, University of Rochester, Rochester, NY 14627, USA}
\author{Kevin Lyons}
\affiliation{Department of Physics and Astronomy, University of Rochester, Rochester, NY 14627, USA}
\affiliation{Center for Coherence and Quantum Optics, University of Rochester, Rochester, NY 14627, USA}
\author{Meiting Song}
\affiliation{The Institute of Optics, University of Rochester, Rochester, New York 14627, USA}
\author{Jaime Cardenas}
\affiliation{The Institute of Optics, University of Rochester, Rochester, New York 14627, USA}
\author{Andrew N. Jordan}
\affiliation{Department of Physics and Astronomy, University of Rochester, Rochester, NY 14627, USA}
\affiliation{Center for Coherence and Quantum Optics, University of Rochester, Rochester, NY 14627, USA}
\affiliation{Institute for Quantum Studies, Chapman University, Orange, CA 92866, USA}
\affiliation{A. N. Jordan Scientific, LLC, 200 Hibiscus Dr. Rochester, New York 14618, USA}

\date{\today}

\begin{abstract}
We present an integrated design to precisely measure optical frequency using weak value amplification with a multi-mode interferometer. The technique involves introducing a weak perturbation to the system and then post-selecting the data in such a way that the signal is amplified without amplifying the technical noise, as has previously been demonstrated in a free-space setup. We demonstrate the advantages of a Bragg grating with two band gaps for obtaining simultaneous, stable high transmission and high dispersion. We numerically model the interferometer in order to demonstrate the amplification effect. The device is shown to have advantages over both the free-space implementation and other methods of measuring optical frequency on a chip, such as an integrated Mach-Zehnder interferometer.
\end{abstract}

\maketitle

\section{Introduction} \label{sec:Intro}

Weak value amplification (WVA)~\cite{aharonov1988result,Duck1989,Dressel2015} can be used to amplify small parameters without amplifying certain types of technical noise, in order to obtain a higher signal-to-noise ratio~\cite{Lyons2018}. The technique has been used to obtain enhanced measurements of a variety of parameters, such as the angular deflection of a mirror~\cite{Dixon2009,Martinez2017}, beam displacements~\cite{Hosten787}, and temperature changes~\cite{Salazar2015}. Here, we propose an integrated interferometer design that uses WVA to sensitively measure changes in optical frequency. 

There are many different types of sensors that have been developed to precisely measure the frequency of a laser~\cite{Dobosz2017,Fox1999,Junttila1990,Yan2010,Vargas2016,Hori1989}. This particular design is inspired by a previous experiment using free space optics~\cite{Starling2010}, but is implemented in an integrated optics environment. The design is a multi-mode Mach-Zehnder interferometer (MZI)~\cite{Steinmetz2019,Song2020} with a small mode perturbation which is coupled to the relative phase between arms. The output power is sorted such that all the information content is concentrated into a small fraction of the light, giving full precision with less light reaching the detector. This allows us to use a much higher input power without saturating the detector. The power that does not reach the detector can be discarded, used as a reference, or recycled~\cite{Dressel2013,Lyons2015}, which could allow for better precision. Using this design, we set a fundamental precision limit of $19~\text{Hz}/\sqrt{\text{Hz}}$ using readily achievable waveguide parameters and $2~\text{mW}$ of detected power ($40~\text{mW}$ of input power), and show that this is better than the precision given by an equivalent standard MZI~\cite{Pezze2007}.

Integrated optics makes a nice platform for miniaturizing interferometers. It has been proposed for making a series of interferometers as needed in boson sampling experiments \cite{crespi2013integrated}, and for single interferometers such as the integrated Sagnac interferometer \cite{menon2003all,jahn1996monolithically}. These integrated interferometers have a wide range of applications, including gyroscopes and accelerometers~\cite{Geen2002,Sorrentino2012,Shaeffer2013,Zandi2010}. Using an integrated optics platform automatically stabilizes the interferometer from drifts in mirror position or air currents, and makes it easier to parallelize the sensing system. Another advantage of working on a chip is that it is possible to implement high-efficiency avalanche photodetectors using nanophotonic techniques \cite{assefa2010reinventing}.The low noise properties of these detectors enable detection of weak optical signals at very high speed.

This paper is organized as follows. In Section~\ref{sec:Int}, we lay out the integrated interferometer design and demonstrate the WVA effect, using theory and numerical simulations. In Section~\ref{sec:Bragg}, we study the dispersive element of the interferometer, and discuss the advantages of using a Bragg grating with two band gaps. In Section~\ref{sec:Prec}, we calculate the precision of the device using Fisher information and quantify the advantage over a MZI. In Section~\ref{sec:Err}, we discuss several common error types that can arise in this kind of interferometer, and show that our WVA design reduces the effect of certain errors compared to a MZI. We conclude in Section~\ref{sec:Con}.

\section{Interferometer design} \label{sec:Int}

\begin{figure}
\includegraphics[scale=0.18]{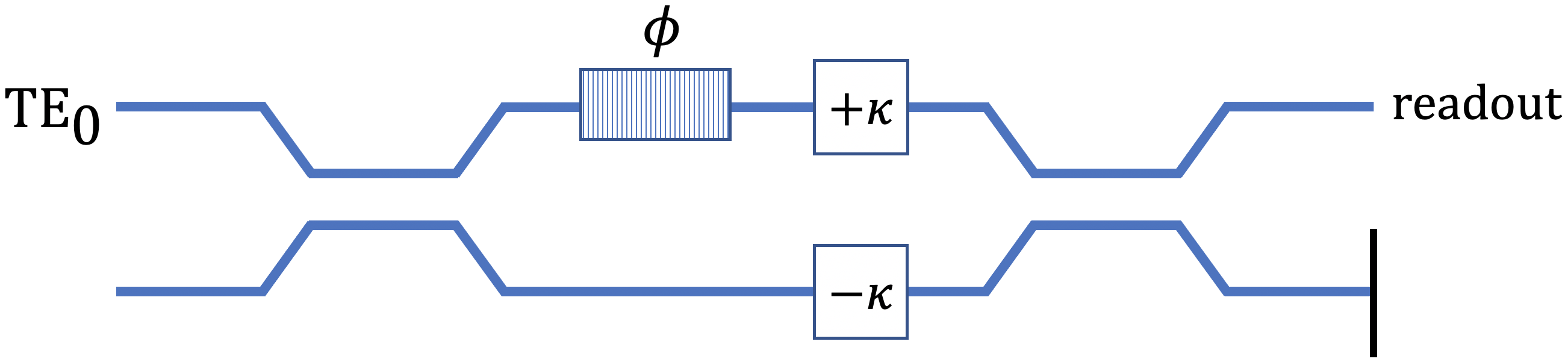}
\caption{The proposed interferometer design. The $\text{TE}_0$ mode enters into the upper arm. A frequency-dependent relative phase of $\phi(\omega)$ is added using a Bragg grating, followed by oppositely tilting phase fronts with wavenumber $\kappa$ using the mode converter design shown in Fig.~\ref{fig:converter}. We choose the amplification factor $\kappa$ such that $\phi(\omega)\ll\kappa\ll 1$. The upper output port is read out using one of the two methods proposed in Section \ref{sec:post}.\label{int}} 
\end{figure}

\subsection{Background}
Weak value amplification consists of three steps: (1) pre-selection, where the system is prepared in an initial state; (2) a weak perturbation to the system state; and (3) post-selection, where the system is projected onto a final state, which is chosen to be nearly orthogonal to the initial state. The result is that only a small fraction of the data is retained, but that small fraction contains nearly the entire information content about the parameter being measured. Consequently, we can use more input power for better precision without saturating the detector.

We consider an infinite planar dielectric waveguide where the core, with index of refraction $n_1$, exists everywhere in space for $|x| \le d$. The cladding layers fill the remainder of space, and for simplicity we assume they both have the same index of refraction $n_2$, where $n_2 < n_1$ so that the wave is guided. We take the $z$-direction to be the direction of propagation and assume $\partial_y U = 0$, where $U$ is any component of the field. We make use of the first two transverse electric modes $\text{TE}_0$ and $\text{TE}_1$ (although a similar analysis can be applied to the $\text{TM}$ modes). The first two $\text{TE}$ modes are
\be
\text{TE}_0(x,z) =e^{i \beta_0 z} \begin{cases}
B_0^- e^{\gamma_0 (x+d)}, & x< -d  \\
A_0 \cos(K_0 x), & -d < x < d \\
B_0^+ e^{-\gamma_0 (x-d)}, & x>d,
\end{cases}
\ee
\be
\text{TE}_1(x,z) =e^{i \beta_1 z} \begin{cases}
B_1^- e^{\gamma_1 (x+d)}, & x< -d \\
A_1 \sin(K_1 x), & -d < x < d \\
B_1^+ e^{-\gamma_1 (x-d)}, & x>d,
\end{cases}
\ee
where $K_{0,1}=\sqrt{n_1^2 k_0^2-\beta_{0,1}^2}$ are transverse wavenumbers, $\gamma_{0,1}=\sqrt{\beta_{0,1}^2-n_2^2k_0^2}$ are decay constants, and $\beta_{0,1}$ are the propagation constants associated with the two modes~\cite{AgrawalBook}. The amplitudes are given by $B_0^\pm = A_0 \cos(K_0d)$ and $B_1^\pm = \pm A_1 \sin(K_1 d)$, where $A_{0,1}$ are determined by normalization. The propagation constants can be determined by solving the transcendental equation
\be \label{eq:transc}
\gamma_{m}d=K_{m}d\tan(K_{m} d-m\pi/2),
\ee 
where $m=0,1,\ldots$ labels the $\text{TE}_m$ modes. For this design, we choose waveguide parameters such that there are only two solutions, corresponding to $m=0,1$. The highest supported guided modes is the number $m$ for which
\be 
k_0 d\sqrt{n_1^2-n_2^2}\geq\frac{m\pi}{2},
\ee 
so we choose parameters such that this inequality is satisfied when $m=1$. For numerical simulations throughout this paper, we choose $\lambda=1550~\text{nm}$, $d=0.3~\mu\text{m}$, and indices of refraction $n_1=1.98$ and $n_2=1.45$, corresponding to a silicon nitride core and silica cladding~\cite{Blumenthal2018}.

\subsection{Pre-selection}
We propose an interferometer design, shown in Fig.~\ref{int}, which uses a small mode perturbation to sensitively read out the optical frequency of the input beam. We start with an injected $\text{TE}_0$ mode in the upper arm, which can be represented in the joint $(\text{path})\otimes(\text{mode})$ space as
\be 
\psi=\begin{pmatrix}
1 \\
0
\end{pmatrix}\otimes\begin{pmatrix}
1 \\
0
\end{pmatrix}
\ee 
where the path vector refers to the two arms of the interferometer in the (upper, lower) basis, and the mode vector is in the ($\text{TE}_0$, $\text{TE}_1$) basis. This state is split with a 50/50 directional coupler, which uses evanescent coupling to transfer power between the two waveguides. The power in the upper and lower waveguides during the directional coupler is
\be 
\begin{split}
I_u&\propto \cos^2(\kappa_c L) \\
I_l&\propto \sin^2(\kappa_c L),
\end{split}
\ee 
where $\kappa_c$ is a coupling constant~\cite{GhatakBook}. To obtain a $50/50$ coupler, we choose the length of the coupler to be $L=\frac{\pi}{2\kappa_c}$, resulting in
\be 
\psi=\frac{1}{\sqrt{2}}
\begin{pmatrix}
1 \\
i
\end{pmatrix}\otimes\begin{pmatrix}
1 \\
0
\end{pmatrix}.
\ee 
In one of the arms, the light encounters a dispersive medium, such as a Bragg grating, that imparts a frequency-dependent phase $\phi(\omega)$. The grating has a medium index that depends sharply on the frequency, resulting in a steep dispersion relation. This is discussed further in Section \ref{sec:Bragg}. The state picks up a relative phase $\phi(\omega)$ between the two arms that depends sensitively on the optical frequency,
\be 
\psi=\frac{1}{\sqrt{2}}
\begin{pmatrix}
e^{i\phi/2} \\
ie^{-i\phi/2}
\end{pmatrix}\otimes\begin{pmatrix}
1 \\
0
\end{pmatrix}.
\ee

\subsection{Mode perturbation}

\begin{figure}
\includegraphics[scale=0.18]{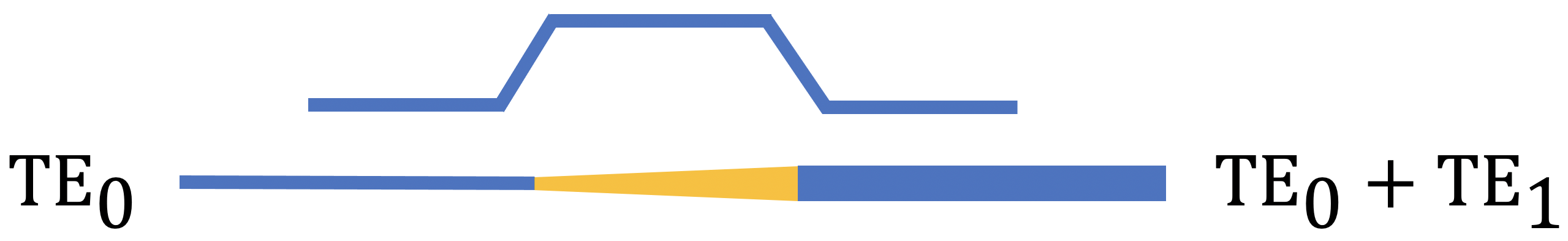}
\caption{Design of a mode converter, which applies a small perturbation to the mode profile. A small fraction $\kappa^2$ of the power is siphoned off into an auxiliary waveguide using a directional coupler. The original waveguide, which only supports the $\text{TE}_0$ mode, enters a tapering region where it gradually increases in width until it supports both $\text{TE}_0$ and $\text{TE}_1$. This occurs slowly enough that light in the $\text{TE}_0$ mode will tend to stay in the $\text{TE}_0$ mode. The final width is chosen such that its $\text{TE}_1$ mode is matched to the $\text{TE}_0$ mode of the auxiliary waveguide. The light from the auxiliary waveguide is then coupled back into the original waveguide.} \label{fig:converter}
\end{figure}

After introducing the relative phase, we apply opposite tilted phase fronts with wavenumber $\kappa$ (where $\kappa$ is taken to be real) by converting a small fraction $\kappa$ of the $\text{TE}_0$ mode into $\text{TE}_1$. The opposite change can be engineered across the other waveguide. We present a design for a ``mode converter'', shown in Fig.~\ref{fig:converter}, in order to perform the transformation
\be 
\text{TE}_0\rightarrow\sqrt{1-\kappa^2}\text{TE}_0 \pm i\kappa\text{TE}_1,
\ee 
where opposite signs are used in each arm. This is similar to the transformation caused by a beam splitter, but in mode space. It is analogous to the beam deflection caused by a prism or mirror tilt in the free space setup~\cite{Starling2010,Dixon2009}. The parameter $\kappa$ should be chosen such that $\phi\ll\kappa\ll 1$ in order to realize the weak value effect. The state after the mode perturbation, written as a non-separable vector in the same $(\text{path})\otimes(\text{mode})$ basis, is
\be \label{eq:balanced}
\psi=\frac{1}{\sqrt{2}}
\begin{pmatrix}
\sqrt{1-\kappa^2}e^{i\phi/2} \\
i\kappa e^{i\phi/2} \\
i\sqrt{1-\kappa^2}e^{-i\phi/2} \\
\kappa e^{-i\phi/2}
\end{pmatrix}.
\ee 

\subsection{Post-selection} \label{sec:post}

The light in the two arms of the interferometer is then combined using another $50/50$ directional coupler. The $\text{TE}_0$ and $\text{TE}_1$ modes have different coupling constants $\kappa_c$, so we must be careful to choose a length where both modes have transferred half their power. After this directional coupler, the state is
\be
\psi=i\begin{pmatrix}
\sqrt{1-\kappa^2}\sin(\phi/2) \\
\kappa\cos(\phi/2) \\
\sqrt{1-\kappa^2}\cos(\phi/2) \\
-\kappa\sin(\phi/2)
\end{pmatrix}.
\ee 
The upper and lower arms have total intensity (normalized relative to the input intensity)
\be 
\begin{split}
I_u&=(1-\kappa^2)\sin^2(\phi/2)+\kappa^2\cos^2(\phi/2)\approx\kappa^2 \\
I_l&=(1-\kappa^2)\cos^2(\phi/2)+\kappa^2\sin^2(\phi/2)\approx 1-\kappa^2,
\end{split}
\ee 
where $\kappa\ll 1$, so we refer to the corresponding output ports as the ``dark port'' and ``bright port'' respectively. We post-select on the dark port, which gives the state
\be \label{eq:dark-unnorm}
\psi_d=i\begin{pmatrix}
\sqrt{1-\kappa^2}\sin(\phi/2) \\
\kappa\cos(\phi/2)
\end{pmatrix}\approx i\begin{pmatrix}
\frac{\phi}{2} \\
\kappa
\end{pmatrix}
\ee 
to first order in the approximations $\phi\ll 1$ and $\kappa\ll 1$. If we are interested in measuring the carrier frequency via the phase, then we can rewrite the mode as
\be \label{eq:dark}
\psi_d \approx i\kappa\left[\text{TE}_1(x) + \left(\frac{\phi}{2\kappa}\right)\text{TE}_0(x)
\right],
\ee
so after renormalization, we have mainly a $\text{TE}_1$ mode with a small amount of $\text{TE}_0$ mode added in. The phase $\phi$ is ``amplified'' by $1/\kappa\gg 1$, and the post-selection probability is given by
$\kappa^2$. 

We suggest two different methods to read out the amplified phase. One method is to measure the ratio between $\text{TE}_0$ and $\text{TE}_1$ modes, giving a signal
\be \label{eq:S1}
S=\frac{\phi}{2\kappa}.
\ee
This mode ratio can be read out using a separate multi-mode interferometer at the dark port which has an output power that depends on the mode ratio. It could also be read out by bringing in a new waveguide, whose fundamental mode frequency is equivalent to the $\text{TE}_1$ mode of the original waveguide, and which will not support the original $\text{TE}_0$ mode. In this way, we can siphon off the $\text{TE}_1$ mode while leaving the $\text{TE}_0$ mode untouched. Photodetectors would then be placed at the ends of those waveguides, and the relative intensity would be read out. Applied to the case of mode (\ref{eq:dark}), one detector would collect only the information signal, and register intensity $(\phi/2)^2$, while the other would measure intensity $\kappa^2$, which would carry no frequency information. The latter signal can be monitored as a reference, discarded, or recycled~\cite{Dressel2013,Lyons2015}. As is shown in Section~\ref{sec:Prec}, measuring the mode ratio is an optimal measurement, i.e. it gives the best possible frequency precision. This is amplified by a factor of $1/\kappa$ when compared with the standard MZI signal, $S_{MZI}\propto\sin\phi\approx\phi$, derived by taking the intensity difference between the two output ports.

\begin{figure}
\includegraphics[scale=0.22] {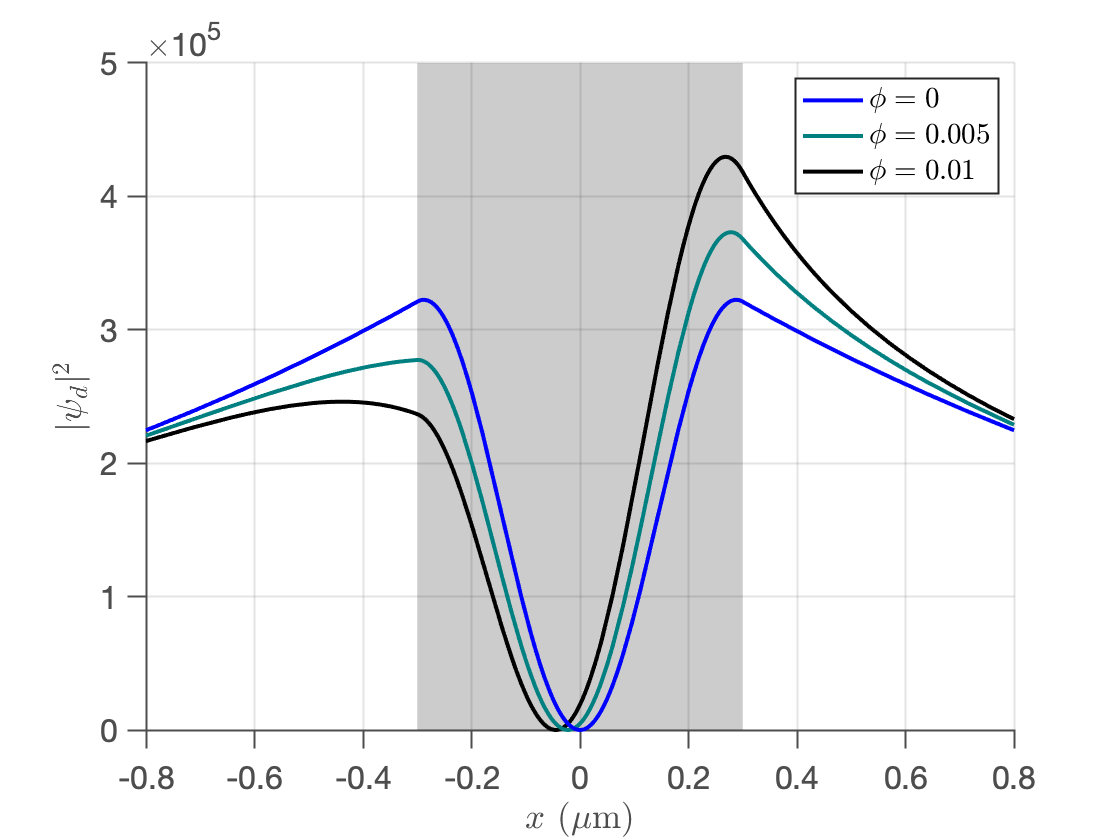}
\caption{The intensity profile in the dark port for a range of relative phases $\phi$. The shaded region is the core of the waveguide. When $\phi=0$, the two lobes are balanced, meaning $I_L=I_R$. As the phase increases, the mode profile shifts to the right, and the magnitude of the signal \eqref{eq:S} grows proportionally to $\phi/\kappa$, where $\kappa=0.05$. \label{fig:modes}}
\end{figure}

A second readout method is to measure the displacement of the mode profile. This is a closer analogy with free space experiments, where the phase is read out by measuring the beam deflection with a split detector. In this case, the expectation value of the transverse position in the dark port depends on the amplified phase,
\be \label{eq:exp-x}
\langle x\rangle \propto \phi(\omega)/\kappa.
\ee
The intensity profile is shown for various values of $\phi$ in Fig.~\ref{fig:modes}. This could be measured using a Y-branch in the waveguide, terminated at the end of the sample by fast photodiodes. The normalized difference signal from the photodiodes in the two arms is
\be \label{eq:S}
S = \frac{I_R - I_L}{I_R+I_L} \approx \frac{\alpha\phi(\omega)}{\kappa},
\ee
where $I_L=\int_{-d}^0dx|\psi(x)|^2$ and $I_R=\int_0^ddx|\psi(x)|^2$ are the intensities in the left and right half of the dark port waveguide, and $\alpha\equiv\int_{-d}^0dx\text{TE}_0\text{TE}_1-\int_0^ddx\text{TE}_0\text{TE}_1$ is a mode constant which does not depend on $\phi$ or $\kappa$. Unlike the mode ratio method, this readout scheme is not optimal, meaning it cannot saturate the ultimate bound on frequency precision (this is discussed further in Section~\ref{sec:Prec}). Also, the true value of $S$ diverges more quickly from the linear approximation we have used, leading to a smaller working range in $\phi$. For these reasons, we will use the mode ratio readout throughout the rest of this paper.

\section{Slow light using a double Bragg grating} \label{sec:Bragg}

\subsection{Bragg grating theory}

\begin{figure}
\centering
\includegraphics[scale=0.2]{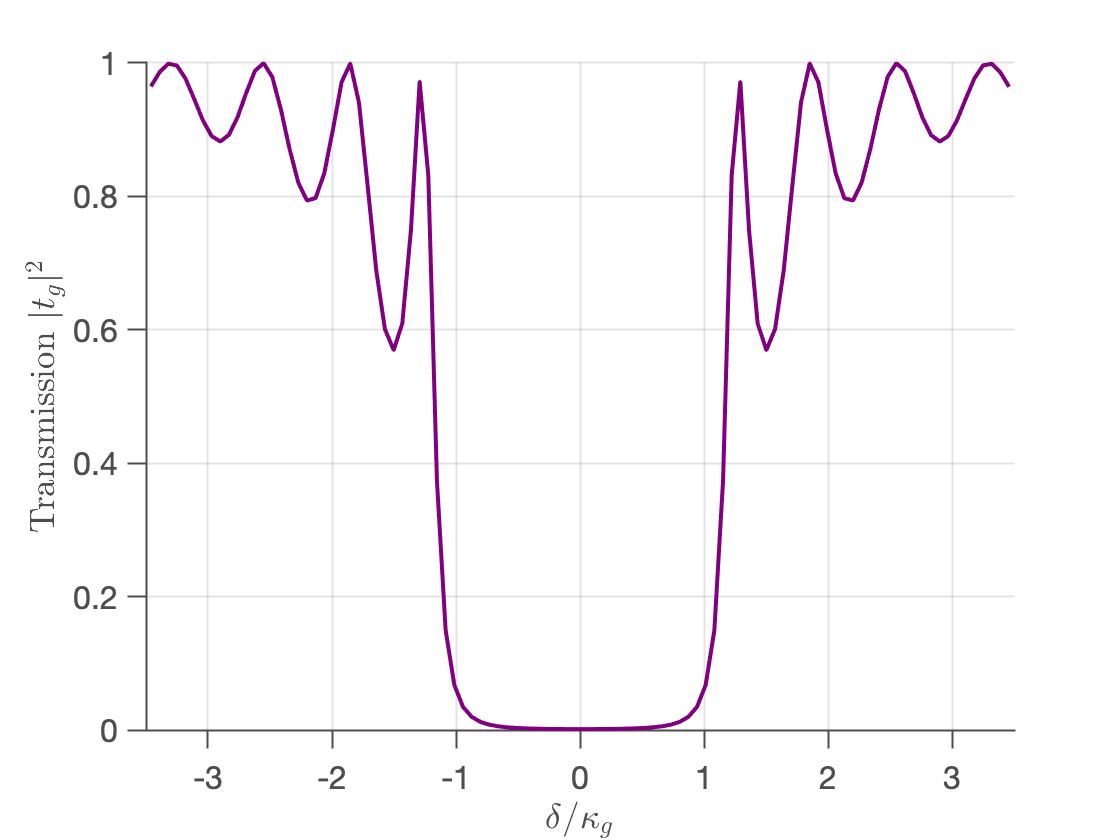}
\includegraphics[scale=0.2]{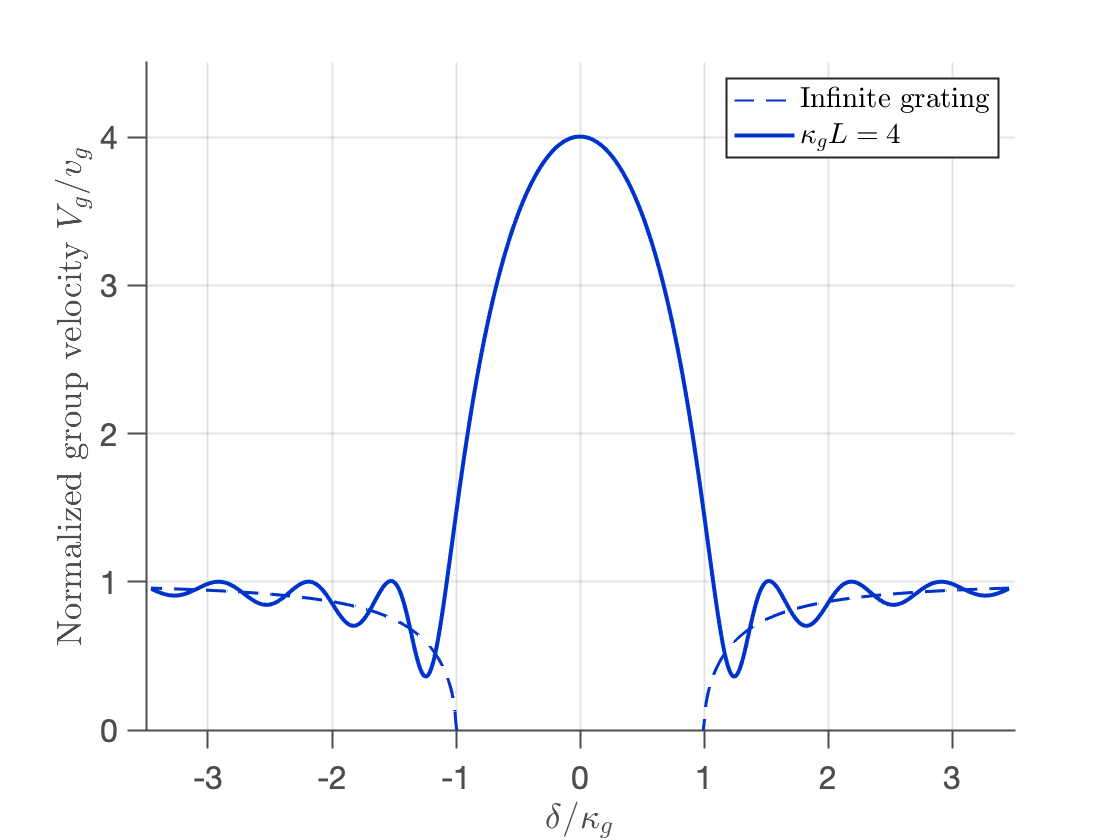}
\caption{(a) Transmission and (b) normalized group velocity of the $\text{TE}_0$ mode in a Bragg grating of length $L=6.58~\text{mm}$ with $\kappa_g L = 4$. The dashed blue curve shows the result of \eqref{eq:Vg-infinite}, which assumes a grating of infinite length. Inside the band gap, the altered wavenumber $q$ is imaginary, so the wave is evanescent and its amplitude decays as it moves through the grating. If the grating is infinite, there is zero transmission throughout the entire band gap. However, if the grating is finite, the evanescent wave can tunnel through the grating with reduced power and superluminal group velocity. This is called the Hartman effect, and has been observed experimentally in fiber Bragg gratings ~\cite{Longhi2001,Longhi2003}. The two curves converge as we move away from the band gap. These plots were created using the fundamental matrix method, which is described in Appendix~\ref{app:bragg}.} \label{fig:gv}
\end{figure}

In order to create strong dispersion, we consider the use of a Bragg grating in one arm of the interferometer~\cite{AgrawalBook,Wen2012}. Another method of creating a frequency-dependent phase is to use the dispersion provided by one or more ring resonators~\cite{Schwelb2004}, but here we restrict the analysis to Bragg gratings. A Bragg grating is a periodic alternating index of refraction
\be
n(z) = \bar{n}+\delta n_g(z),
\ee
where $\bar{n}=\frac{\beta\lambda}{2\pi}$ is the effective index of refraction of the waveguide for the $\text{TE}_0$ mode (the only mode traveling through the grating), and $\delta n_g(z)$ has spatial periodicity $\Lambda$. The periodic grating opens a photonic band gap, where certain wavenumbers cannot propagate through the grating, centered at $\lambda=2\bar{n}\Lambda$. This is analogous to the conduction band gap in semiconductors. The traveling waves exhibit dispersion, and slow light effects can appear near the band gap. This is related to the nonvanishing first derivative of the index of refraction with respect to frequency. While any periodic index of refraction is sufficient to produce this effect, we will focus here on the simplest case of a sinusoidal grating, $\delta n_g = n_a \cos(2 \pi z/\Lambda)$, which can be created using laser etching from an interference pattern as one fabrication technique.

For a Bragg grating of infinite length, the propagating field takes the form $e^{\pm i q z}$, where
\be
q = \pm \sqrt{\delta^2 - \kappa_g^2}
\label{dispersion}
\ee
is the new wavenumber, $\delta(\omega) = \frac{n_1}{c}(\omega-\omega_B) \equiv \beta(\omega) - \beta_B$ is the detuning of the wavenumber from the Bragg wavenumber $\beta_B = \pi/\Lambda$, and $\kappa_g = \pi n_a/\lambda$ is the coupling coefficient between the forward and backwards modes. If the detuning $|\delta|$ is less than the coupling $\kappa_g$, there is no (real) solution, and the traveling wave mode cannot exist. Outside this band, the wave number is modified by the grating to become $\beta_e = \beta_B \pm q$. The dependence of $q$ on $\omega$ indicates the presence of dispersive effects. We expand $\beta_e$ in a Taylor series in $\omega$,
\be
\beta_e(\omega) = \beta_0^g + (\omega - \omega_0) \beta_1^g + \frac{1}{2} (\omega - \omega_0)^2 \beta_2^g + \ldots
\ee
and study the first derivative, which sets the group velocity of the grating (or sensitivity of the phase to frequency),
\be \label{eq:Vg-infinite}
V_g = 1/\beta_1^g = \pm v_g \sqrt{1-\kappa_g^2/\delta^2},
\ee
where $v_g = \left(\partial \delta / \partial \omega\right)^{-1} = \left(\partial \beta/ \partial \omega\right)^{-1}$ is the native group velocity of the waveguide. As $|\delta|$ approaches $\kappa_g$, the group velocity slows to zero, while for $|\delta| \gg \kappa_g$, the grating is irrelevant. 

The group velocity of a finite-length grating can be obtained using
\be \label{eq:Vg-finite}
\frac{L}{V_g}=\frac{\partial \arg{(r_g)}}{\partial \omega},
\ee
where 
\be
r_g=\frac{i\kappa_g\sin(qL)}{q\cos(qL)-i\delta\sin(qL)}
\ee 
is the complex reflection coefficient~\cite{Poladian1997, PetermannThesis}. This takes into account edge effects, which cause oscillations about the expression given in \eqref{eq:Vg-infinite}. Equation \eqref{eq:Vg-finite} converges to \eqref{eq:Vg-infinite} as the length of the grating becomes very large. A comparison is shown in Fig.~\ref{fig:gv}.

While it may be tempting to work very close to $|\delta| = \kappa_g$ to get the slowest light, this is generally a bad idea because a large fraction of the light is reflected. Fig.~\ref{fig:gv} demonstrates the inconvenient conclusion that when the group velocity is low, the transmission is also low. If we move a bit away from $\delta = \kappa_g$ and take the length of the grating to be fairly long, we can still obtain high transmission while keeping relatively low group velocity.

One of the awkward features of working near the band gap is that while there is dispersion that will help in our frequency measurement, the transmission is changing (typically rapidly) as frequency (or detuning) is changed, as shown in Fig.~\ref{fig:gv}. We would like to have relatively constant transmission throughout the working frequency range, while also having small group velocity, or high dispersion.

\subsection{Two resonances}

\begin{figure}
\centering
\includegraphics[scale=0.2]
{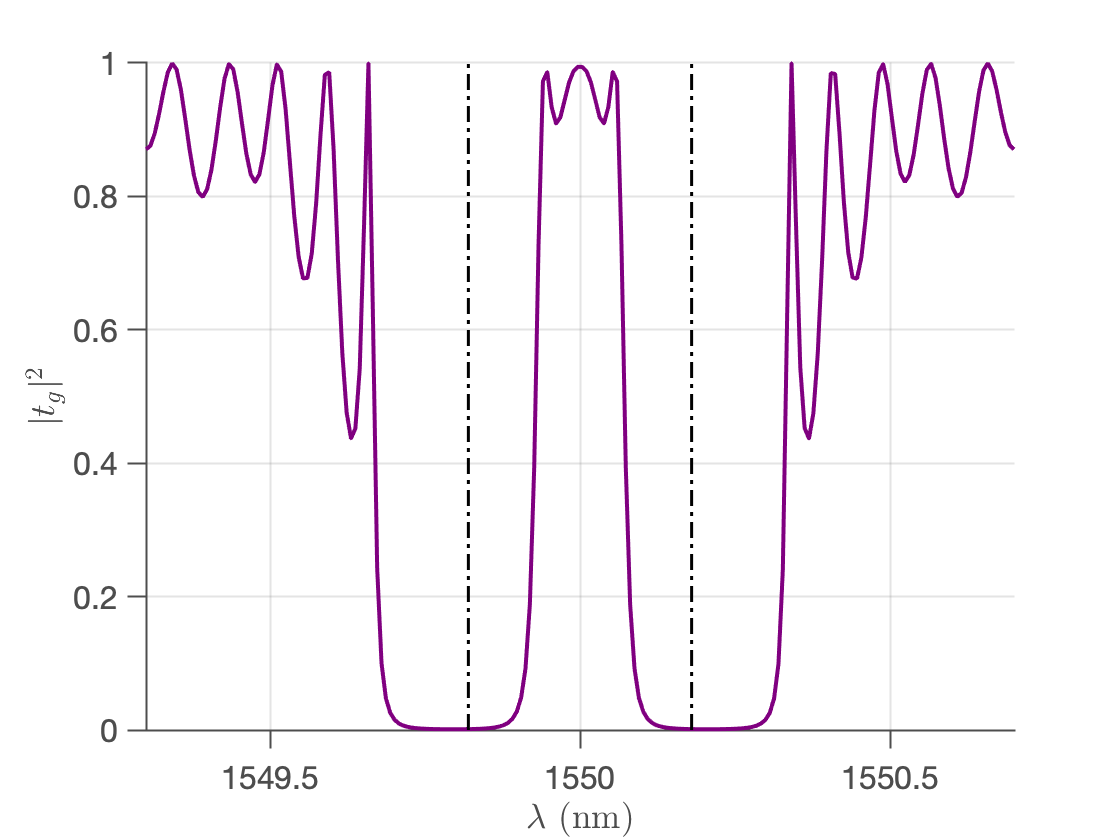}
\includegraphics[scale=0.2]
{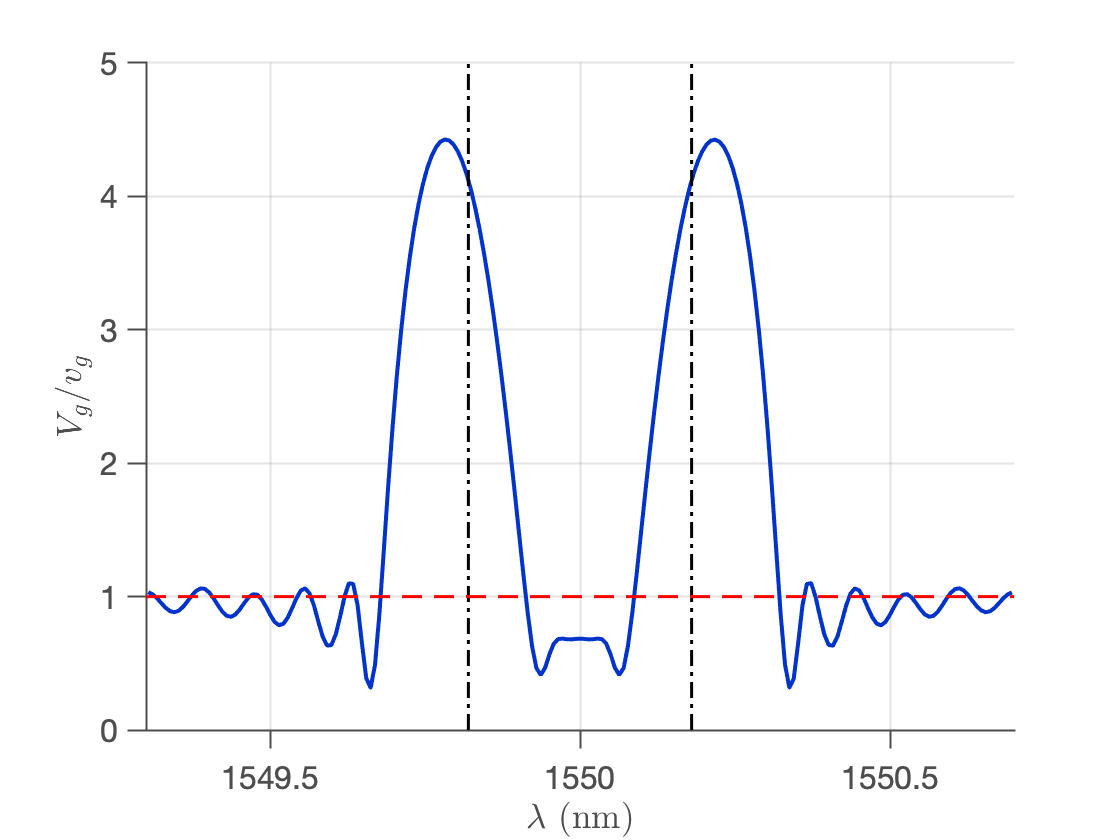}
\includegraphics[scale=0.2]
{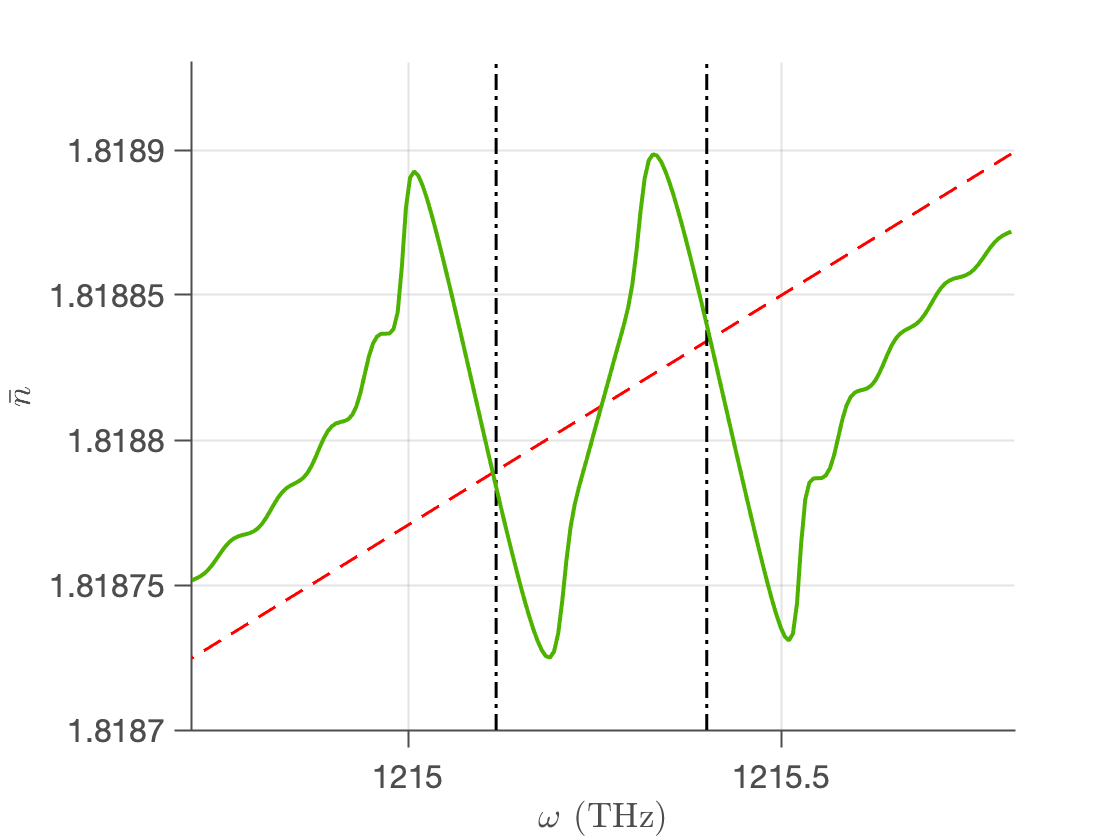}
\caption{Numerical simulation of the double Bragg grating proposal. Transmission, group velocity (normalized with respect to the native group velocity of the waveguide), and effective index of refraction are plotted for a double grating with band gaps centered at $\Lambda_1=1549.82~\text{nm}$ and $\Lambda_2=1550.18~\text{nm}$ (shown by the black dash-dotted lines), and $n_a=n_b=3\times 10^{-4}$. In between the band gaps, there is relatively stable high transmission ($\sim 0.99$) and low group velocity ($\sim 68\%$ of the native group velocity). This also corresponds to a large derivative $\frac{\partial \bar{n}}{\partial\omega}$. These plots were created using the thin layer method, which is described in Appendix~\ref{app:bragg}. \label{fig:doublegrating}}
\end{figure}

One way to mitigate this difficulty that was proposed in the context of atomic resonances is to work in the region between two resonances \cite{Camacho2006,starling2012double}. This technique allows high-precision frequency measurements with relatively high optical transmission so as to minimize the optical losses. We can use an analogous idea here by having a double periodicity in the grating, 
\be
\delta n(z) = n_a \cos(2 \pi z/\Lambda_1) + n_b \cos(2 \pi z/\Lambda_2).
\ee
This grating will then open up two photonic band gaps centered at $\beta_1 = \pi/\Lambda_1$ and $\beta_2 = \pi/\Lambda_2$, each with a width given by $\kappa_{g1,2}$, where $\kappa_{g1} = \pi n_a/\lambda_1$ and $\kappa_{g2} = \pi n_b/\lambda_2$. By arranging a region of parameter space that allows propagating modes between $\beta_1 + \kappa_{g1}$ and $\beta_2 - \kappa_{g2}$, we can accomplish the same basic physics that was accomplished in the atomic system: a region of frequency space that has fairly high transmission, but also high dispersion (very slow light). A numerically simulated comparison of transmission and group velocity for a grating with two band gaps is shown in Fig.~\ref{fig:doublegrating}. As expected, there are two band gaps centered at $\Lambda_1$ and $\Lambda_2$. The regions inside the band gaps exhibit superluminal group velocity but close to zero transmission. On the outside of the two band gaps, the transmission and group velocity both oscillate. In the center of the two band gaps, there is high transmission and relatively low group velocity, which are the desired qualities of the double Bragg grating concept design. The region between the band gaps is small (in Fig.~\ref{fig:doublegrating} the window is $0.36~\text{nm}$) which limits the working frequency range, but the whole region has high transmission and low group velocity.

\section{Sensitivity and precision analysis} \label{sec:Prec}

\subsection{Sensitivity}
The frequency sensitivity of the device is given by
\be 
\Delta S=\frac{\partial S}{\partial\phi}\frac{\partial\phi}{\partial\omega}\Delta\omega,
\ee
where $S$ is the output signal given by \eqref{eq:S1} or \eqref{eq:S} depending on the readout method, $\frac{\partial\phi}{\partial\omega}$ is the dispersion from the Bragg grating, $\Delta\omega$ is the frequency shift to be measured, and $\frac{\partial S}{\partial \phi}=\frac{1}{2\kappa}$ is the amplification factor using the mode ratio readout method. The higher the dispersion and amplification of the device, the more sensitive the signal will be to changes in frequency. The relative phase caused by the Bragg grating, assuming equal path lengths, is given by $\phi=(\beta_e-\beta)L$, so the dispersion is given by
\be \label{eq:dispersion}
\frac{\partial\phi}{\partial\omega}=\frac{L}{v_g}\left(\frac{v_g}{V_g}-1\right).
\ee 
Using $L=6.58~\text{mm}$ (so that $\kappa_g L\approx 4$), $V_g=0.68v_g$, and $v_g\approx0.5c$ (which was obtained numerically using the chosen waveguide parameters), the overall sensitivity is
\be 
\Delta S=(2.5\times10^{4})\frac{1}{2\kappa}\frac{\Delta\omega}{\omega}.
\ee 
For example, if $\kappa=0.05$, this is $\Delta S=(2.5\times10^{5})\frac{\Delta\omega}{\omega}$. Compare this to a MZI, which has $\frac{\partial S}{\partial\phi}\approx1$, resulting in a sensitivity of $\Delta S=(2.5\times10^{4})\frac{\Delta\omega}{\omega}$. The WVA interferometer results in a sensitivity that is enhanced by the amplification factor $\frac{1}{2\kappa}$.

\subsection{Precision}

The Cramér-Rao bound (CRB) gives the fundamental limit on the precision of a parameter being estimated~\cite{Paris2009}. In this case, it gives the minimum detectable change in frequency,
\be 
\Delta\omega^2\geq\frac{1}{\mathcal{F}(\omega)},
\ee 
where $\mathcal{F}(\omega)$ is the Fisher information,
\be 
\mathcal{F}(\omega) = P_0(\partial_\omega \ln P_0)^2 + P_1(\partial_\omega \ln P_1)^2,
\ee 
and $P_0$ and $P_1$ are the probabilities of the two measurement outcomes. For a standard MZI~\cite{Pezze2007}, a photon can arrive at one of two output ports, with probabilities $P_0=\sin^2(\phi/2)$ and $P_1=\cos^2(\phi/2)$, resulting in the minimum detectable frequency change
\be \label{eq:CRB}
\Delta\omega\geq\frac{1}{\sqrt{N}\partial_\omega\phi(\omega)}
\ee 
for $N$ total input photons. The frequency sensitivity depends on the input power and the dispersion $\partial_\omega\phi(\omega)$. For an input power of $2~\text{mW}$, this gives a precision bound of $390~\text{Hz}/\sqrt{\text{Hz}}$. This value depends heavily on the dispersion provided by the Bragg grating.

For the WVA interferometer, we only look at the dark port (which detects no photons when $\phi=0$), and use the mode ratio to make a measurement. The probabilities associated with the $\text{TE}_{0}$ and $\text{TE}_{1}$ modes are $P_0\approx\frac{\phi(\omega)^2}{4}$ and $P_1\approx\kappa^2$. The Cramér-Rao bound in this case is the same as for the MZI, despite only using a small fraction of the available power. The low detection probability is balanced by the amplification of $\phi$, which concentrates the Fisher information into the subset of data being measured. The advantage of the WVA interferometer lies in the fact that we can greatly increase the number of photons without overloading the detector, since most of the input light never reaches the detector. If we increase $N\rightarrow N/\kappa^2$, the minimum detectable frequency change is
\be 
\Delta\omega\geq \frac{\kappa}{\sqrt{N}\partial_\omega\phi(\omega)},
\ee 
which is decreased by a factor of $\kappa\ll 1$. We have increased the amount of input power to achieve better precision, but critically, the amount of power arriving at the detector is the same as it is for a MZI. With $\kappa=0.05$, we can use $20$ times as much input power, so the precision bound is $19~\text{Hz}/\sqrt{\text{Hz}}$.

\subsection{Quantum Fisher information}

When doing parameter estimation using a quantum state, the quantum Fisher information (QFI) $\mathcal{I}$ gives an upper bound on the Fisher information optimized over all possible measurement schemes~\cite{Paris2009}. The QFI for the output of the WVA interferometer, including both ports, is $\mathcal{I}=(\partial_\omega\phi)^2$. This is the maximum possible information that can be gained from the readout. This information is almost entirely concentrated into the dark port, which can be seen by calculating the QFI for the dark port only, $\mathcal{I}\approx(\partial_\omega\phi)^2(1-\kappa^2)$. As $\kappa\rightarrow 0$, this approaches the total QFI contained in the system. This means in the $\kappa\ll 1$ limit, the WVA interferometer channels almost all the information in the system into the dark port, even though it contains only a small fraction of the input power.

An optimal measurement will be one whose Fisher information saturates the QFI. If we consider the mode ratio readout method, the Fisher information is $(\partial_\omega\phi)^2$, as in \eqref{eq:CRB}. In this case $\mathcal{F}=\mathcal{I}$, so reading out the mode ratio is an optimal measurement. If instead we read out the mode displacement as in \eqref{eq:S}, we can treat $P_{0,1}=I_{L,R}$ and take $\phi\rightarrow 0$, to obtain the Fisher information
\be 
\mathcal{F}=(\partial_\omega\phi)^2\frac{4(\int_0^ddx\text{TE}_0\text{TE}_1)^2}{\int_{-d}^ddx|\text{TE}_1|^2}<(\partial_\omega\phi)^2.
\ee 
In this case $\mathcal{F}<\mathcal{I}$, which means this is not an optimal measurement scheme. Using the chosen waveguide parameters, $\mathcal{F}\sim 0.6\mathcal{I}$, which results in a precision loss by a factor of only $\sqrt{0.6}$ compared to the mode ratio method, so it could still be worth using if it is easier to implement experimentally.

\section{Error analysis} \label{sec:Err}

Some of the main errors and sensitivities that affect interferometers are: (1) bias offset error, (2) bias instability, (3) temperature sensitivity, and (4) shock and vibration sensitivity~\cite{Pupo2016,Pachwicewicz2018}. Bias offset error is the systematic error that the instrument shows when it is at rest. Typically this constant meter reading is a product of fabrication and can be subtracted off in the calibration process. The bias instability, however, corresponds to a relatively slow random walk in the bias offset, which leads to slow random error, which will limit the accuracy of the sensor. In the subsections that follow, we make an analysis and simulation of errors (1), (2), and (3). Error (4) is difficult to simulate because the error source is not the system itself, but rather the detector electronics and aspects outside the interferometer. It is therefore more accurately measured in the experimental testing of the device. Also, the integrated optical readout greatly reduces shock and vibration sensitivity compared to the free space version. Unless the integrated optical semiconductor structure itself is shattered or severely deformed, this will not be relevant. 

We expect that WVA techniques will help in three ways. First, the systematic bias offset in the meter reading will be suppressed compared to a MZI, provided that source of the drift is outside the system itself. Second, even if the offset drifts, as expected in the bias instability noise induced by the detector and surrounding environment, the WVA will also suppress it, even if it is unknown. Third, and perhaps most importantly, the fact that standard commercial detectors saturate after a few tens of mW of power allows us to use much higher input power, and even though the detectors measures only a fraction of that power, we attain a precision equal to that of the entire input power, giving an important practical advantage.

\subsection{Bias offset errors}

\begin{figure}
    \centering
    \includegraphics[scale=0.22]{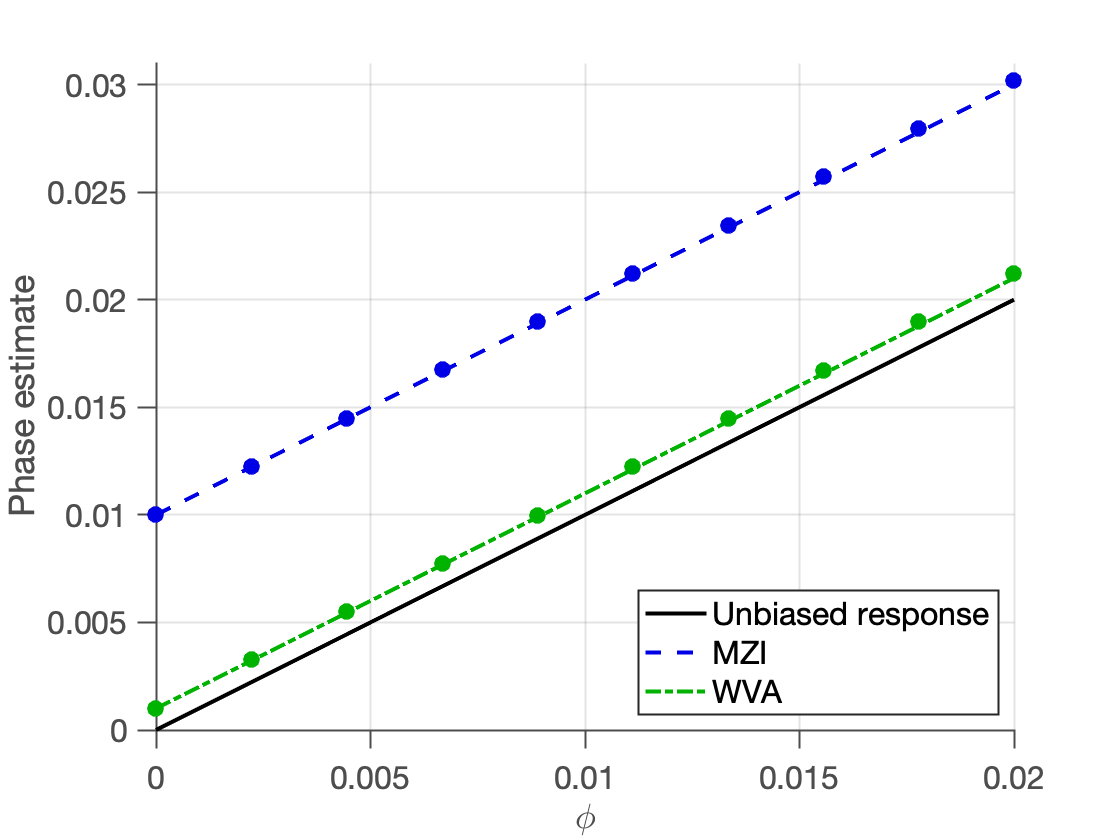}
    \caption{The response of the phase readout in the MZI and WVA interferometer with an applied bias of $b=0.01$ (i.e. $1\%$ of photon detections are misread) and amplification factor $\kappa=0.05$. The closer the response is to the unbiased response (solid black), the more accurate the phase estimate. The offset in the WVA readout is suppressed by a factor of $2\kappa=0.1$ compared to the MZI readout. The points on the plot are numerically simulated results, and the lines are the theoretical result $\phi+\phi_0$. The simulated results begin to deviate slightly from the theory as $\phi$ increases since the actual signal is nonlinear, but this is not an issue when $\phi\ll\kappa\ll 1$.}
    \label{fig:bias-offset}
\end{figure}

The phase readout can be biased if there are slight imperfections in the interferometer components. This bias could also come from slightly imbalanced loss in the two waveguides leading to the detector, or any number of other possible asymmetries in the system. In practice, this means the device will detect a frequency that is different from the true frequency by some constant offset. In the WVA interferometer, the frequency readout is amplified while the bias offset error is not, so the error is suppressed~\cite{Pang2016}. 

We can model a bias offset error as a certain fraction of photons being misread by the detector. In a MZI, where the signal is given by the difference between the lower and upper waveguide intensities $S=I_l-I_u$, we add in a bias by supposing the detector misreads a fraction $b$ of the upper waveguide photons as being in the lower waveguide, $S=I_l+bI_u-(1-b)I_u$. This results in the phase estimate being $\phi\rightarrow\phi+\phi_0$, where the phase offset is $\phi_0\approx b$. In the WVA interferometer, the signal is given by the mode ratio in the dark port, so we similarly define the bias offset error to be a fraction $b$ of photons in $\text{TE}_0$ being detected as if they were in $\text{TE}_1$. This results in a signal $S\approx\frac{\phi}{2\kappa}+b$, where we have assumed $b\ll 1$ to get a linear offset. The phase offset in this case is $\phi_0\approx 2\kappa b$, which has been suppressed by a factor of $2\kappa$. The suppression of a bias offset error by the WVA effect is shown in Fig.~\ref{fig:bias-offset}.

The offset discussed here is easy to subtract off and calibrate away, so long as it does not drift in time. We will discuss the case of the drifting offset in the next subsection. The interferometer can be calibrated by measuring the phase offset when $\phi=0$, and then accounting for this bias in any future measurements.

\subsection{Bias instability}
\begin{figure}[h]
    \centering
    \includegraphics[scale=0.22]{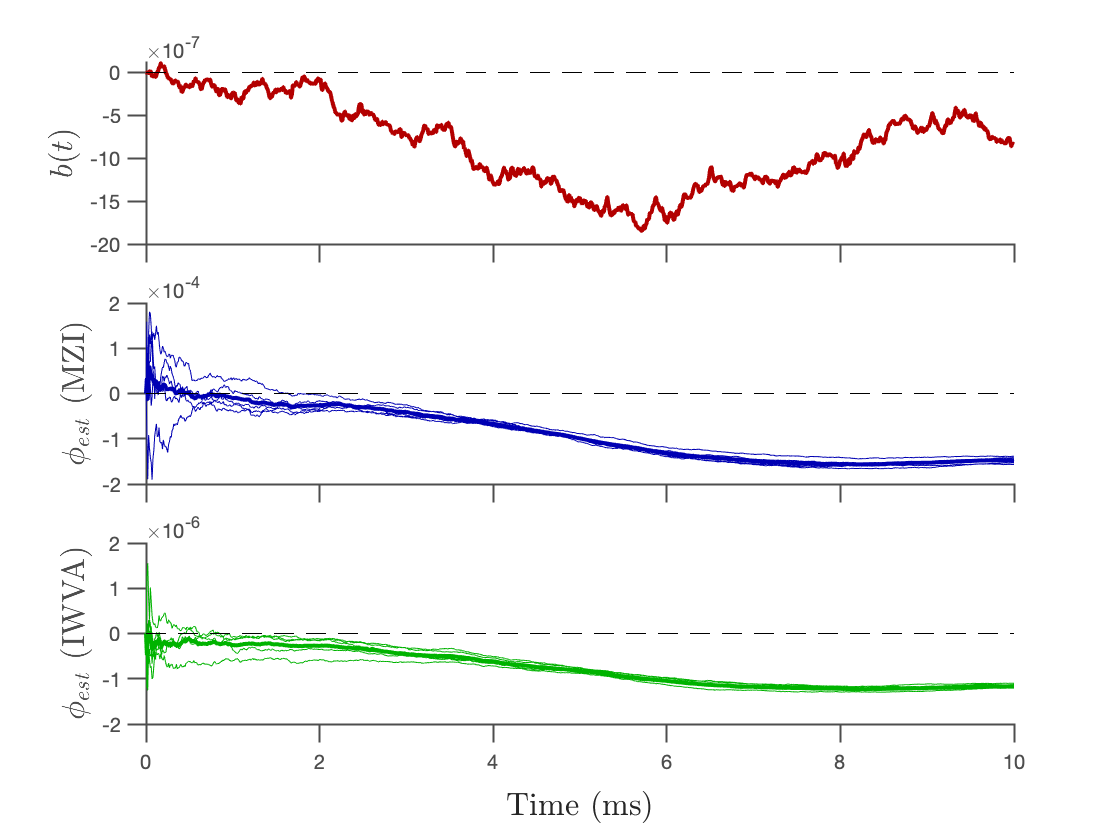}
    \caption{The phase estimate $\phi_{est}$ being read out by the detector is shown under the influence of a time-varying bias $b(t)$ with a mean of zero and standard deviation of $1\times10^{-5}\text{/s}$. The actual phase is $\phi=0$. We use $2~\text{mW}$ of power at a wavelength of $1550~\text{nm}$. We plot $1000$ intermediate time steps, so each step includes $\sim10^{11}$ photons, each of which is detected as being in one of two output ports. Five sample trajectories (thin lines) and their average (thick line) are shown for both WVA and MZI under the same time-dependent bias. The bias instability $b(t)$ can be seen in the slope of the sum of signals $\sim \phi_{est}t$. For the MZI, we treat each photon as arriving either in the upper or lower output port. For the WVA interferometer, we treat each photon as if it arrived at the most likely position $\left<x_{L,R}\right>$ on the left or right side of the dark port, and use \eqref{eq:exp-x}. The time drift for the WVA interferometer is suppressed by around two orders of magnitude compared to the MZI (note the different $y$-axis scales). This advantage is quantified more precisely in Fig.~\ref{fig:Allan}.}
    \label{fig:bias-inst-ARW}
\end{figure}

\begin{figure}[h]
    \centering
    \includegraphics[scale=0.22]{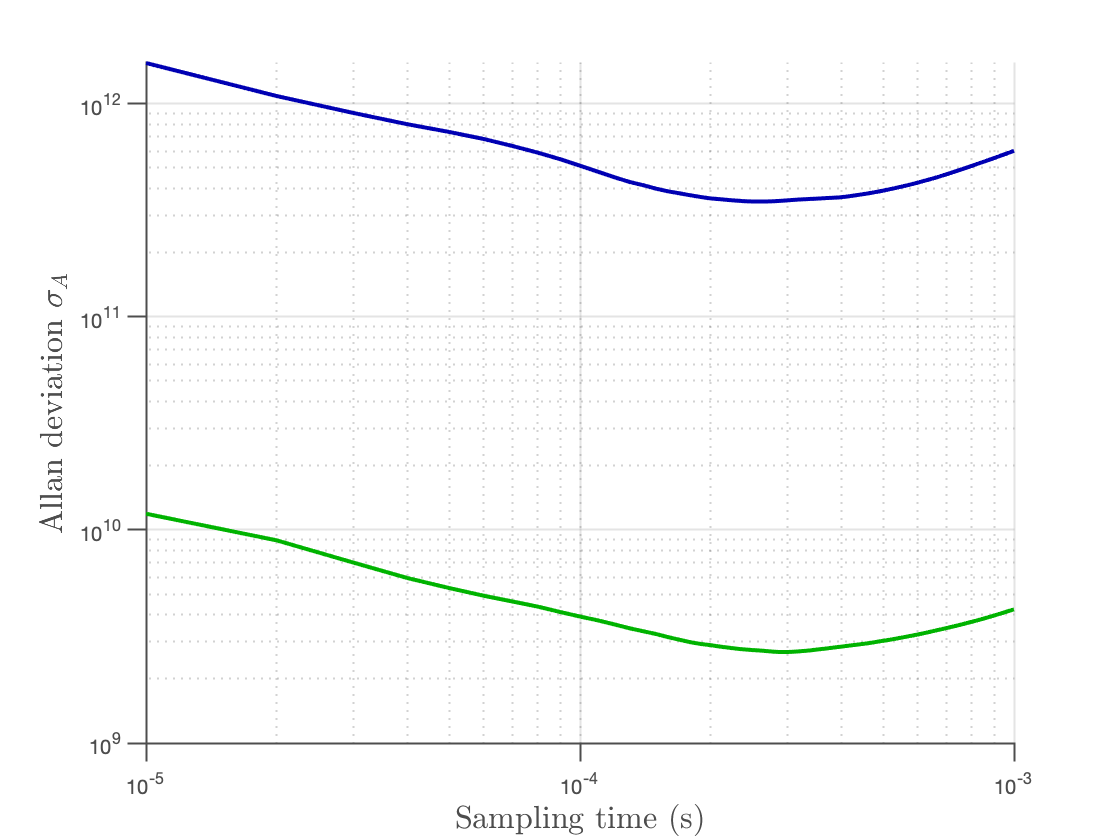}
    \caption{A comparison of the Allan deviation for MZI and WVA, over a range of sampling times $\tau$, for the average detector signal shown in Fig.~\ref{fig:bias-inst-ARW}. The values on the $y$-axis are large because $\sigma_A(\tau)$ is based on the cumulative signal, which is proportional to the number of detected photons. The Allan deviation is larger at first because of high-frequency noise, reaches a minimum as the sampling time smoothes out this noise, then increases again to due to long-term drifts in $b(t)$. The Allan deviation in WVA, with an amplification factor of $\kappa=0.05$, is lowered by an average factor of $\sim 0.0076$. This means this sensor is affected by the bias instability less by that same factor than the MZI. The suppression is not exactly the same for all $\tau$ since this is based on an average of only $5$ trajectories, as shown in Fig.~\ref{fig:bias-inst-ARW}. The advantage of using WVA grows if we use a smaller $\kappa$.}
    \label{fig:Allan}
\end{figure}
Bias in the interferometer is more difficult to deal with when it varies in time. If the bias factor $b(t)$ is time-dependent, the readout signal $S$, and therefore the phase estimate $\phi$, will depend on the time-averaged bias factor $\left< b(t) \right>$. We suppose that we can calibrate away the constant offset from the fabrication of the device, and only have to contend with the time-varying bias instability. 
We model the bias instability as a random walk $b=\sum_i B_i$ in the bias offset, where $B_i$ is a normally distributed random variable with mean $\langle B_i\rangle=0$ and standard deviation $1\times10^{-5}/s$. This number is typical of what we expect from thermal drifts.


The drifting bias offset $b(t)$ causes a corresponding drift in the phase offset $\phi_0(t)$, and we know from the previous section that $\phi_0=b$ for the MZI and $\phi_0=2\kappa b$ for WVA. The standard deviation of the drift in phase offset for the same amplitude and time scale of the bias instability is then $5.7 \times 10^{-4} \ ^\circ /\rm s$ for the MZI, and is suppressed to $5.7 \times 10^{-5} \ ^\circ /\rm s$ for the WVA with $\kappa=0.05$. The sum of detected signals for a MZI and the WVA interferometer (using the mode displacement readout technique for closer analogy with the MZI signal) under influence of such a bias offset is shown in Fig.~\ref{fig:bias-inst-ARW}. As with the constant bias offset, this suppression can be enhanced by increasing the amplification in the WVA interferometer.

To quantify the advantage of this method, we compute the Allan variance of the MZI and the WVA interferometer~\cite{Pupo2016,Pachwicewicz2018,riley2008handbook}. The Allan variance is a measure of how quickly the rate of an accumulating signal is changing. It is calculated by grouping a sequence of $N$ data points $x_i$ into time bins of length $\tau=m dt$, and using
\be 
\sigma_A^2=\frac{1}{2\tau^2(N-2m)}\sum_{i=1}^{N-2m}(x_{i+2m}-2x_{i+m}+x_i)^2.
\ee 
The Allan variance is generally minimized over the sampling time $\tau$. If the bias is very unstable, the signal will change by very different amounts over each sampling time, and the Allan variance (and Allan deviation $\sigma_A$) will be large. A comparison of the Allan deviation for MZI and WVA is in Fig.~\ref{fig:Allan}, which confirms that the effect of bias instability is greatly reduced by the WVA effect.

\subsection{Thermal effects}
\begin{figure}
    \centering
    \includegraphics[scale=0.22]{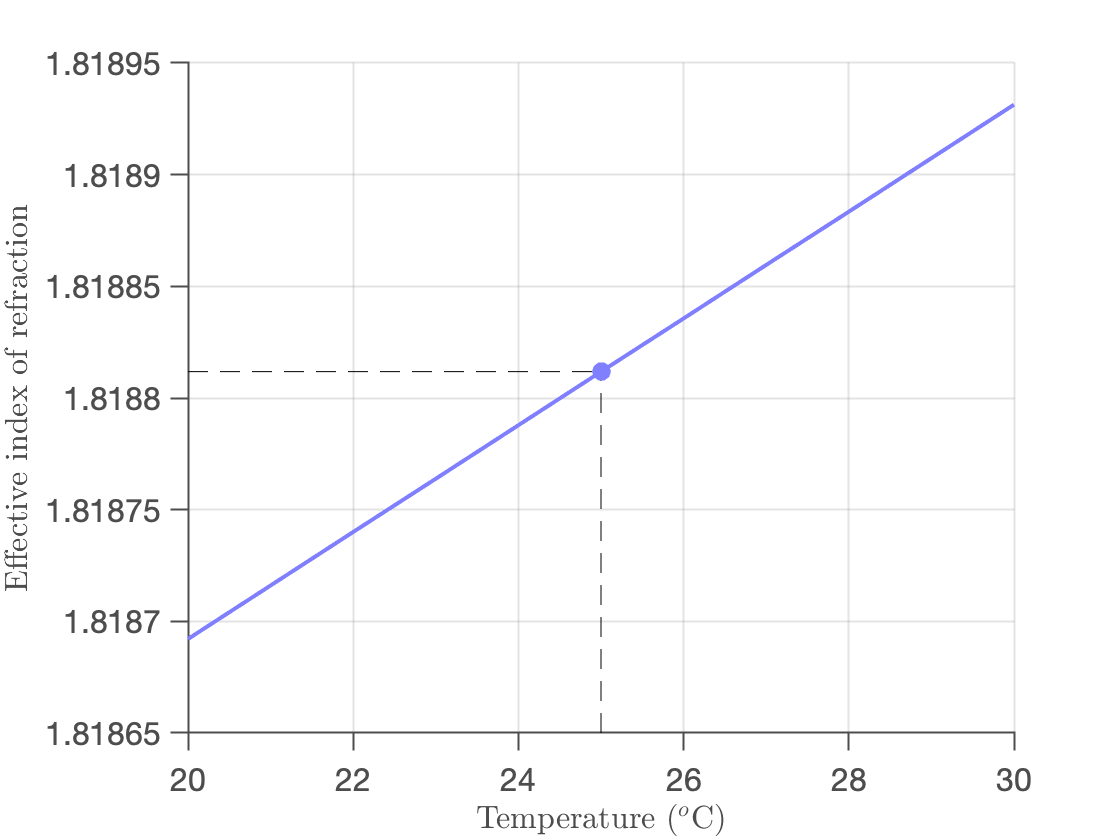}
    \caption{Numerical simulation of the effective index of refraction $\bar{n}$ as the temperature changes, by solving \eqref{eq:transc} for the appropriate range of $n_1$ and $n_2$. This uses the thermo-optic coefficients $\frac{dn}{dT}=2.45\times10^{-5}$ for the Si$_3$N$_4$ core and $\frac{dn}{dT}=9.5\times10^{-6}$ for the SiO$_2$ cladding. The slope gives an effective thermo-optic coefficient of $\frac{\partial\bar{n}}{\partial T}=2.39\times 10^{-5}$, which is between $\frac{\partial n_1}{\partial T}$ and $\frac{\partial n_2}{\partial T}$ as we should expect. The value of $\bar{n}$ is highlighted at $T=25^\text{o}\text{C}$, corresponding to the values $n_1=1.98$ and $n_2=1.45$ used elsewhere in this paper.}
    \label{fig:temp}
\end{figure}

The temperature stability of the sensor is affected by the thermally dependent index of refraction of the waveguide materials. On one hand, this is helpful because heaters can be placed near the waveguides in order to fine tune their optical properties. On the other hand, undesired temperature shifts can create systematic errors in the accuracy of the measurement readings. We quantify this behavior using the thermo-optic effect in our materials of choice (Si$_3$N$_4$ for the core and SiO$_2$ for the cladding),
\be
\begin{split}
\frac{dn_1}{dT} &= 2.45 \times 10^{-5}/ ^\circ {\rm C} \\
\frac{dn_2}{dT} &= 9.5 \times 10^{-6}/ ^\circ {\rm C},
\end{split}
\ee
where $T$ is the temperature~\cite{arbabi2013measurements}. The propagation of the $\text{TE}_0$ mode through the waveguide is described by the effective index of refraction $\bar{n}=1.82$, which was determined numerically using the chosen waveguide parameters and the techniques laid out in Section~\ref{sec:Int}. The effective thermo-optic coefficient is determined in Fig.~\ref{fig:temp} to be $\frac{\partial\bar{n}_0}{\partial T}=2.39\times 10^{-5}$. The same argument also applies to $\text{TE}_1$, which will have a different effective thermo-optic coefficient, $\frac{\partial\bar{n}_1}{\partial T}=1.19\times 10^{-5}$. The most sensitive temperature dependence will be in the acquired phase in the system. The phase difference between the $\text{TE}_0$ mode in the two arms (assuming a uniform shift of $n$) will be given by
\be
\Delta \phi = 2\pi \bar{n}(T) \Delta L/\lambda,
\ee
where $\Delta L$ is the path length difference between arms of the interferomter. Consequently, a temperature fluctuation $\delta T$ will result in a phase drift of
\be \label{eq:phase-drift}
\delta \phi = \frac{\partial \phi}{\partial T} \delta T,
\ee
where 
\be
\frac{\partial \phi}{\partial T} = \frac{2\pi \Delta L}{\lambda} \frac{\partial\bar{n}_0}{\partial T} \approx  0.001/^\circ {\rm C}, \label{phasederivative}
\ee
where we have used a path length difference of $10~\mu\text{m}$ and wavelength of $1550~\text{nm}$. In experiments, this length mismatch is likely an overestimate, so we may well have better temperature robustness in actual experiments. Using the dispersion from the Bragg grating given in \eqref{eq:dispersion}, we can convert this phase sensitivity into a frequency sensitivity of
\be
\frac{1}{\omega}\frac{d\omega}{dT} = \frac{1}{\omega}\frac{d\omega}{d\phi}\frac{d\phi}{dT} \approx 3.9\times 10^{-8}/^\circ \text{C}.
\ee

Since the modes have different thermo-optic coefficients, there will also be some unwanted phase difference
\be 
\delta\phi_{01}=\frac{\partial\phi_{01}}{\partial T}\delta T
\ee 
between $\text{TE}_0$ and $\text{TE}_1$ that accumulates between the mode converter and the post-selection, where
\be 
\frac{\partial \phi_{01}}{\partial T} = \frac{2\pi \Delta L}{\lambda}\left( \frac{\partial\bar{n}_0}{\partial T}-\frac{\partial\bar{n}_1}{\partial T}\right) \approx  5\times10^{-4}/^\circ {\rm C}.
\ee 
This relative phase between modes causes the dark port mode in \eqref{eq:dark} to be altered to
\be 
\psi_d\approx i\kappa\left[\text{TE}_1(x)+e^{i\delta\phi_{01}}\left(\frac{\phi}{2\kappa}\right)\text{TE}_0(x)\right].
\ee 
If the readout is done using the mode ratio (see Section~\ref{sec:Int}), then the signal is unaffected. However, if the readout is done using the displacement of the mode profile, then the amplification in \eqref{eq:S} is reduced when the modes are out of phase, such that
\be 
S\approx\frac{\alpha\phi(\omega)}{\kappa}\cos\delta\phi_{01}.
\ee 
This is a negligible effect compared to the linear phase drift in \eqref{phasederivative}, since $\delta\phi,\delta\phi_{01}\ll 1$ at reasonable temperature drifts.

\subsection{Shock and vibration sensitivity}
One of the principal advantages of using an integrated optics chip is stability in the presence of shocks and vibrations (see e.g. \cite{monovoukas_integrated_2000}). The field boundary conditions and relative phases are preserved under a translation of the device, and should represent an extremely small contribution to measurement uncertainty for this device. Unlike sensors based on mechanical elements, the optical readout inside the integrated geometry makes this system highly robust to shocks to the system. The only possible damage to the system is if the shock is so great that the chip itself becomes mechanically damaged. Another possible weakness of the system is the process of coupling light into and out of the system, and directing it into a detector. This can be overcome in several ways. Further enhancements to the robustness can be obtained if necessary by integrating the photodetectors into the chip geometry.

\section{Conclusions} \label{sec:Con}
This analysis outlines a path forward to implementing a chip-scale frequency sensor enhanced using weak value amplification. This is done with a dispersive element to imprint a frequency-sensitive phase shift on the light, a mode converter to introduce a small perturbation, and an interferometer to read out that phase via intensity differences on two detectors. We have shown that the WVA method concentrates the total information content into a small fraction of the input power, which allows us to amplify the signal without amplifying the technical noise. We esimate a precision bound of $19~\text{Hz}/\sqrt{\text{Hz}}$ based on readily attainable waveguide parameters and $2~\text{mW}$ of detected power, compared to $390~\text{Hz}/\sqrt{\text{Hz}}$ for a standard MZI. This technique also mitigates errors due to a constant or time-varying bias in the interferometer. While the waveguide geometry necessitates many differences from the free space version, we have shown how the effect can be realized, and provided a number of options for doing so, including an analysis of Bragg gratings with either one or two resonances. The integrated optics environment offers several advantages compared to the free space version, including easier parallelization and resistance to shock and vibration sensitivity.

There are several directions that can be taken to enhance this method. Nonclassical light, such as a squeezed state, could be injected into the unused input port of the interferometer so as to apply existing quantum metrology methods to further improve phase sensitivity beyond the standard quantum limit. Recycling of the bright port photons could also be implemented by adding a guided loop to re-inject lost light. The basic interferometer design provided here can also be applied to many other problems in metrology. By replacing the dispersive element with other components, it is possible to encode other small parameters into the relative phase, which can then be amplified in the same way.

\section{Acknowledgments}
We are grateful to Marco Lopez and John C. Howell for helpful comments and discussions. This work was funded by Leonardo DRS technologies.

\section{Disclosures}
ANJ discloses that a portion of this research was conducted outside of the University of Rochester through his LLC. Financial interests include ownership and fiduciary roles in the LLC.

\appendix

\section{Bragg grating simulations} \label{app:bragg}

The following section details the numerical methods used to create the Bragg grating simulations in Section~\ref{sec:Bragg}. The grating is treated as a lumped element, and its effects are calculated using a transfer matrix method. Two variants will be discussed here: the fundamental transfer matrix approach and the thin layer approach.

\subsection{Fundamental matrix approach}

The fundamental matrix approach is computationally inexpensive and easy to implement, but only works for a single-frequency grating. It consists of finding a matrix $\mathbf{F}$ that relates the forward- and backward-traveling waves $A_f$ and $A_b$ at either side of the grating. This can be written as a matrix equation,
\be
\begin{bmatrix}
A_f (0)\\
A_b (0)
\end{bmatrix}
=
\begin{bmatrix}
F_{11} & F_{12} \\
F_{21} & F_{22}
\end{bmatrix}
\begin{bmatrix}
A_f (L) e^{-i\beta L}\\
A_b (L) e^{i\beta L}
\end{bmatrix},
\ee
where we set $A_b(L)$ to zero. The matrix elements of $\mathbf{F}$, which can be calculated using coupled-mode theory \cite{AgrawalBook,PetermannThesis}, are
\be
\begin{split}
F_{11} &= F_{22}^* = \left(\cos{qL} + \frac{i\delta}{q}\sin{qL}\right) e^{i\beta_B L} \\
F_{21} &= F_{12}^* = -\frac{i\kappa_g}{q}\sin{qL} e^{i(\beta_B L - \pi/2)}.
\end{split}
\ee
The reflection and transmission coefficients are then
\be \label{eq:tg-rg}
\begin{split}
r_g &= \frac{F_{21}}{F_{11}} \\
t_g &= \frac{1}{F_{11}},
\end{split}
\ee
and can be used to calculate the transmission and group delay of the grating using \eqref{eq:Vg-finite}.

\subsection{Thin layer approach}

\begin{figure}[h]
\includegraphics[scale=0.45]{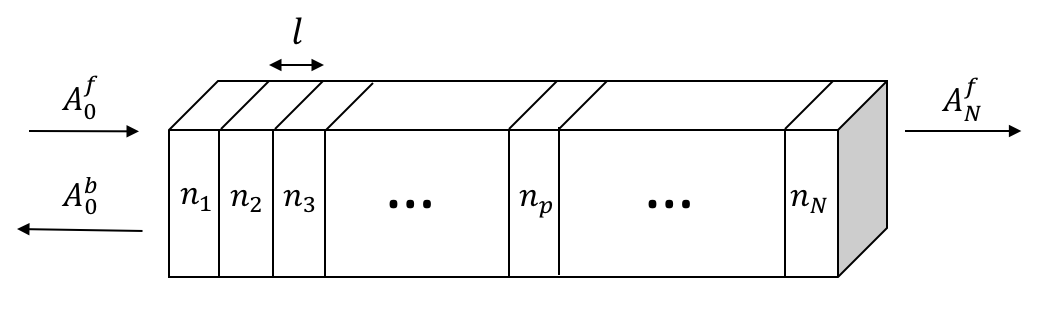}
\caption{A model of a Bragg grating with arbitrary $n(z)$ used in the thin layer approach. \label{fig:thinlayer}}
\end{figure}

To simulate a grating with two band gaps, or any other arbitrary $n(z)$, we can use the thin layer method \cite{Muriel1997,PetermannThesis}. We model the grating as $N$ thin segments of length $l\ll L$, each with an approximately constant index of refraction, as shown in Fig.~\ref{fig:thinlayer}. The index of segment $p$ is $n_p \equiv n(pl)$. We can relate the forward- and backward-traveling waves across an interface using 
\be
\begin{bmatrix}
A^f_{p-1}\\
A^b_{p-1}
\end{bmatrix}
=
\frac{1}{2n_p}\mathbf{M}_p \mathbf{T}_p
\begin{bmatrix}
A^f_p\\
A^b_p
\end{bmatrix},
\ee
where
\be 
\frac{1}{2n_p}\mathbf{M}_p = 
\frac{1}{2n_p}\begin{bmatrix}
n_{p-1}+n_{p} & n_{p-1}-n_{p} \\
n_{p-1}-n_{p} & n_{p-1}+n_{p}
\end{bmatrix}
\ee
is derived \cite{PetermannThesis} using the Fresnel equations to relate the forward- and backward-traveling waves at each interface, and 
\be 
\mathbf{T}_p = 
\begin{bmatrix}
e^{-i\frac{2\pi}{\lambda}n_p l} & 0 \\
0 & e^{i\frac{2\pi}{\lambda}n_p l}
\end{bmatrix}.
\ee
implements the phase resulting from propagation by a distance $l$ through a segment with index of refraction $n_p$. The full effect of the grating is the product of the matrices representing each layer,
\be
\mathbf{F} = \prod_{p=1}^N \frac{1}{2n_p}\mathbf{M}_p\mathbf{T}_p.
\ee
The transmission and reflection coefficients can then be calculated in the same way as before, using \eqref{eq:tg-rg}. The thin layer approach is more computationally expensive than the fundamental matrix approach since it must include many segments within each grating period in order to be accurate, but it allows for any arbitrary periodic index variation $n(z)$.

\bigskip

\bibliography{refs}

\end{document}